\documentstyle[twocolumn,seceq,epsf]{jpsj}

\newfont{\msam}{msam10}
\newfont{\ninemsam}{msam9}
\newcommand{\lsim}{\mbox{\msam\symbol{46}}}
\newcommand{\ninelsim}{\mbox{\ninemsam\symbol{46}}}
\newcommand{\gsim}{\mbox{\msam\symbol{38}}}

\newcommand{\cavo}{\mbox{$\rm CaV_2O_5$}}
\newcommand{\srcuo}{\mbox{$\rm SrCu_2O_3$}}

\def\runtitle{Spin Excitations in the Two-leg Ladder Antiferromagnet
CaV$_{2}$O$_{5}$}
\def\runauthor{Tetsuo {\sc Ohama}, Masahiko {\sc Isobe} and Yutaka
{\sc Ueda}}

\title{Spin Excitations in the Two-leg Ladder Antiferromagnet CaV$_{2}$O$_{5}$}
\author{Tetsuo {\sc Ohama}$^{1,2,}$\footnote{
Present Address: Department of Physics, Faculty of Science, Chiba University,
Chiba, 263-8522.},
Masahiko {\sc Isobe}$^{1}$ and Yutaka {\sc Ueda}$^{1}$} 
\inst{
$^{1}$Institute for Solid State Physics, University of Tokyo,
Kashiwanoha, Kashiwa, 277-6581\\
$^{2}$Department of Physics, Faculty of Science, Chiba University,
Inage-ku, Chiba, 263-8522
}

\recdate{\hspace{20mm}}

\abst{
The nuclear spin-lattice relaxation rate  at the V sites 
in \cavo\ was measured.
The relaxation rate ($1/T_1$) in the two-leg ladder 
was calculated with a strong-coupling expansion ($J_\perp \gg J_\|$)
and the result was compared with the experimental data.
The comparison of the temperature dependence shows a good agreement and
provides an estimate of both $J_\perp$ and $J_\|$ as well as
the energy gap $\Delta$.
The magnetic field dependence of $1/T_1$ is also reproduced with this
approximation in the  temperature range $T\ \ninelsim\ \Delta /4$.
It is found that the magnon-magnon collision, which leads to a diffusive
behavior, is negligible in this temperature range.
The effect of inter-ladder coupling is investigated
and found to be weak in \cavo .
Thus the strong-coupling expansion calculation
agrees satisfactorily with the experimental data, 
but a discrepancy between the exchange parameters deduced from the magnetic
susceptibility and $1/T_1$ is found.
}

\kword{CaV$_{\sf 2}$O$_{\sf 5}$, NMR, spin ladder, spin diffusion,
nuclear spin relaxation}

\begin{document}\sloppy\maketitle

\section{Introduction}
The discovery of high-$T_{\rm C}$ superconductivity in doped
two-dimensional  antiferromagnets renewed interest
in quantum antiferromagnets.
In particular,
spin-$\frac{1}{2}$ two-leg ladder antiferromagnets have been extensively
studied,
being motivated by a theoretical prediction of the quantum disordered
ground state with an energy gap and possible superconductivity.\cite{Dagotto}
Actually, the energy gap was experimentally observed in \srcuo\cite{Azuma}.
Moreover, superconductivity was discovered in
Sr$_{14-x}$Ca$_x$Cu$_{24}$O$_{41}$ under high pressure.\cite{Uehara}

The low-energy excitations in such gapful spin systems
have not been fully understood yet.
One problem is a significant discrepancy between
the energy gaps determined from the magnetic susceptibility
and the nuclear spin-lattice relaxation rate $1/T_1$.
For the two-leg ladder antiferromagnet \srcuo,
$1/T_1$ suggests a much larger gap $\Delta\sim$
700 K than $\Delta\sim$ 420 K by the susceptibility.\cite{Azuma,Ishida}
Itoh {\it et al.} surveyed the data of the susceptibility and $1/T_1$ of many
gapful quantum paramagnets and pointed out that this discrepancy is observed
widely in one-dimensional gapful systems such as Haldane-gap and two-leg
ladder antiferromagnets.\cite{YItoh}
Kishine and Fukuyama calculated $1/T_1$ with the Majorana fermion
representation of the two-leg ladders.\cite{Kishine}
They found a dominant contribution with three times larger energy gap
than the lowest excitation and conclude that this is responsible for
the discrepancy.
Sachdev and Damle developed a semiclassical continuum theory for 
one-dimensional gapful systems and obtained diffusive behavior in the
long-time spin correlation.\cite{Sachdev97,SachdevQPT}
They showed that $1/T_1$ is not proportional to ${\rm e}^{-\Delta/T}$
but ${\rm e}^{-3\Delta/2T}$ at sufficient but not too low temperatures.
The diffusive behavior was actually observed in magnetic field dependence
of $1/T_1$ in the Haldane-gap antiferromagnet AgVP$_{2}$S$_{6}$\cite{TakiAgV}
and another gapful system (VO)$_2$P$_2$O$_7$.\cite{Furu,Kikuchi}
These theoretical arguments suggest that the intrinsic energy gap is unique
and that the discrepancy was observed because the data fitting was done
at too high temperatures.

The discrepancy between the energy gaps was also observed\cite{Iwase}
in another two-leg ladder antiferromagnet \cavo .\cite{Bouloux, Onoda}
$1/T_1$ at the V sites indicates activated temperature dependence 
with an energy gap of 616 K.
On the other hand,
somewhat smaller gap $\sim$ 460 K was suggested from the NMR shift,
although the data accuracy was insufficient.
\begin{figure}[bt]
  \begin{center}
\epsfxsize=60mm \epsfbox{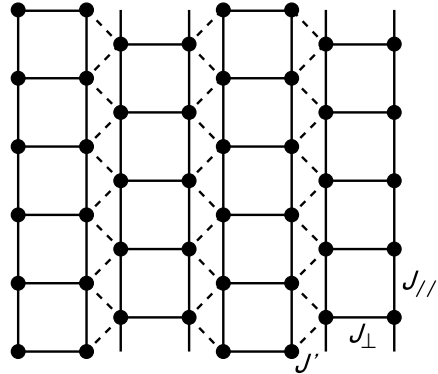}
  \end{center}
\caption{Schematic drawing of dominant exchange interactions in \cavo :
solid lines  for intra-ladder couplings $J_{\parallel}$ and $J_{\perp}$, and
dashed lines for inter-ladder coupling $J'$.}
\label{fig:ex}
\end{figure}

In this compound,
one $3d$ electron per V site occupies the $d_{xy}$-like orbital
in a VO$_{5}$ pyramid,\cite{OhamaCa}
which forms V$_2$O$_5$ layers.
Dominant exchange interactions between electronic spins
at the V sites are schematically shown in Fig.~\ref{fig:ex}:
two exchange interactions, $J_{\parallel}$ and $J_{\perp}$,
between corner-sharing VO$_5$ pyramids form
two-leg ladders and the other between edge-sharing pyramids is frustrating
inter-ladder coupling $J'$.
These exchange interactions form a trellis lattice
similarly to \srcuo.
Onoda and Nishiguchi reported that the temperature dependence of
the magnetic susceptibility
can be fitted well by the isolated dimer model
with the intra-dimer coupling $\sim$ 664 K,\cite{Onoda}
suggesting anisotropic exchange interactions near the strong-coupling limit
($J_\perp \gg J_\|$).
Miyahara {\it et al.} compared the experimental data of the susceptibility 
to quantum Monte Carlo simulations for the trellis lattice.\cite{Miya}
They also concluded that \cavo\ is near the isolated dimers:
the intra-ladder exchange along the rung is dominant
$J_\perp\sim 670$ K, and the others are in the range of
0 $< J_\| <$ 200 K, and $J_\|+J' \sim $ 110 K.
An {\it ab initio} calculation of the exchange parameters was performed
using the LDA + {\it U} method.\cite{Korotin99, Korotin00}
Their result that $J_\perp\sim 608$ K, $J_\|\sim$ 122 K, and $J' \sim -28$ K
agrees with the quantum Monte Carlo simulation.

We report in this article $1/T_1$ measurements at the V sites
in \cavo\ with a powder sample.
The dependence of 1/$T_1$ on the temperature and magnetic field was measured
in a wide temperature range
to compare it to theories in detail.
We calculate $1/T_1$ for the two-leg ladder with the strong-coupling expansion
in \S 3 and compare the result with the experimental data in \S 4.
We estimate the exchange parameters from the temperature dependence.
Next we compare the magnetic field dependence and 
investigate whether the diffusive behavior is observed in \cavo .
Also we examine whether the magnetic susceptibility can be explained 
consistently with the exchange parameters deduced from the analysis of $1/T_1$.

\section{Experimental}

The powder sample used in this study is the same as the previous
NMR shift measurement.~\cite{OhamaCa}
It was magnetically aligned and was found
 by an X-ray diffraction measurement that the alignment field
($H_{\rm al}$) was distributed in the $a$-$b$ plane and mainly along
the $b$ axis.
\begin{figure}[tb]
  \begin{center}
\epsfxsize=85mm \epsfbox{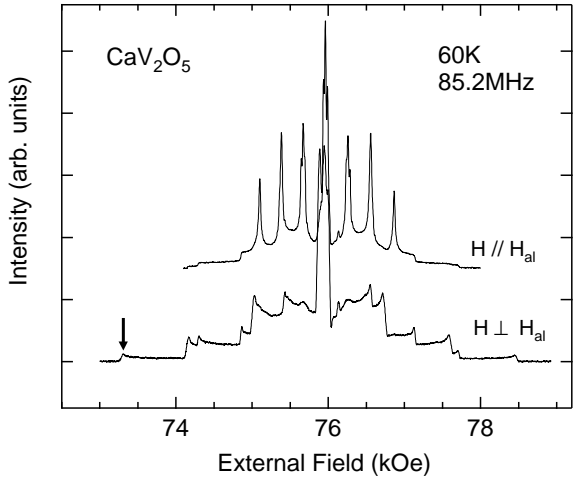}
  \end{center}
\caption{$^{51}$V NMR spectra at 60 K. Satellite transition used in $1/T_1$
measurements is shown by arrow.}
\label{fig:spec}
\end{figure}

We measured $1/T_1$ by observing the
spin-echo recovery of the third satellite transition after an inversion pulse
with  the magnetic field ($H$)
along the maximum principal axis of the electrical
field gradient at the V sites.
At the magnetic field of this satellite transition,
shown in  $^{51}$V NMR spectra\cite{OhamaCa} in Fig.\ref{fig:spec},
the observed nuclear magnetization recovery can be fit well
by the analytical formula of the nuclear relaxation
for the uniaxial electrical field gradient ($\eta=0$)
without cross relaxation between the nuclear Zeeman levels\cite{Narath}
\begin{eqnarray}
\frac{M(t)}{M_{0}}&=& 1-C\left[\frac{1}{84}{\rm e}^{-t/T_{\rm 1}}+
\frac{3}{28}{\rm e}^{-3t/T_{\rm 1}}\right.\nonumber\\
&&+\frac{3}{11}{\rm e}^{-6t/T_{\rm 1}}
+\frac{25}{77}{\rm e}^{-10t/T_{\rm 1}}+\frac{75}{364}{\rm e}^{-15t/T_{\rm 1}}
\nonumber\\
&&\left.+\frac{3}{44}
{\rm e}^{-21t/T_{\rm 1}}+\frac{4}{429}{\rm e}^{-28t/T_{\rm 1}}\right].
\end{eqnarray}
We thereby obtain $1/T_1$ with good accuracy.
At the magnetic field of the other satellite transition,
the spin-echo signal contains signals of other satellite
transition or for other magnetic field directions,
thus the precise measurement of $1/T_1$ is not achievable.

\section{1/$\mib T_{\mib 1}$ in Strong-coupling Two-leg Ladders}

In this section we calculate $1/T_1$
in two-leg ladders with a strong-coupling expansion
to analyze the experimental data.
Sagi and Affleck recently developed a theory of nuclear relaxation
in the Haldane-gap antiferromagnet,\cite{Sagi}
which has an excitation spectrum similar to the anisotropic two-leg
ladder near the strong-coupling limit.
We here calculate $1/T_1$ in the two-leg ladder
following their arguments.
Difference in the hyperfine interaction between the Haldane-gap
and two-leg ladder antiferromanets, that is, the coupling of nuclear spins
to the excitations is essential.
Further, we also calculate $1/T_1$ for a trellis lattice 
including the inter-ladder coupling.

In general, $1/T_1$ due to the fluctuating hyperfine
field $\mib H^{\rm hf}(t)$ is given by
\begin{equation}
\frac{1}{T_1}=\frac{\gamma^2}{2}\int_{-\infty}^{\infty}dt\,\langle\{
H_+^{\rm hf}(t)H_-^{\rm hf}(0)\}\rangle_{\beta}{\rm e}^{i\omega_{\rm N}t}
\end{equation}
where $\gamma$ is the nuclear gyromagnetic ratio, $\omega_{\rm N}\equiv\gamma
 H$, $\{AB\}\equiv(AB+BA)/2$, and $\langle\ldots\rangle_\beta$ is the thermal
average.\cite{Moriya}
Here we take the $z$ axis along the magnetic field.
For the magnetic nuclear sites, we can neglect the off-diagonal part of 
the hyperfine interaction as
\begin{equation}
  {\cal H}_{\rm hf} = \frac{A_\perp}{2}(I_+S_-+I_-S_+)+A_\parallel I_zS_z,
\end{equation}
then $1/T_1$ is
\begin{eqnarray}
\frac{1}{T_1}&=&\frac{A_\perp^2}{2\hbar^2}\int_{-\infty}^{\infty}dt\,\langle\{
S_+(t)S_-(0)\}\rangle_{\beta}{\rm e}^{i\omega_{\rm N}t}.
\end{eqnarray}

For the two-leg ladder antiferromagnet,
the low-lying excitations consist of triplet ``magnons''.\cite{Barnes,Troyer}
The main nuclear relaxation process is the scattering of a thermally excited
magnon by a nuclear spin (the Raman process).
These magnons are not real bosons, since one cannot excite two magnons
on the same rung at the same time,
but at low temperatures $T \ll \Delta$ where the number of excited
magnons is small, we can consider the magnons as independent bosons.
Then $1/T_1$ is
\begin{eqnarray}
\frac{1}{T_1}&=&\frac{A_\perp^2}{2\hbar^2}
\sum_{k,k'}\sum_{\sigma,\sigma'}|\langle k',\sigma'|S_+|k,\sigma\rangle|^2
\left(\frac{\textrm{e}^{-\beta\epsilon_{k\sigma}}+\textrm{e}^{-\beta
\epsilon_{k'\sigma'}}}{2}\right)
\nonumber\\
&&\times\int_{-\infty}^{\infty}dt\,{\rm e}^{it(
\epsilon_{k\sigma}-\epsilon_{k'\sigma'}
-\hbar\omega_{\rm N})/\hbar},
\end{eqnarray}
where $|k\sigma\rangle$ is a one-magnon state with momentum $k$ and $z$
component of spin $\sigma=-1,0,1$, and 
$\beta\equiv 1/T$.

We first consider the two-leg ladder
\begin{equation}\label{eq:ladder}
  {\cal H} = \sum_{r=1}^{N}J_\perp {\mib S}^{\rm L}_r\cdot {\mib S}^{\rm R}_r
+\sum_{r=1}
^{N}J_\|({\mib S}^{\rm L}_r\cdot {\mib S}^{\rm L}_{r+1}+{\mib S}^{\rm R}_r
\cdot{\mib S}^{\rm R}_{r+1}).
\end{equation}
in the strong-coupling expansion;\cite{Barnes,Reigrotzki}
${\mib S}^{\rm L}_r$ and ${\mib S}^{\rm R}_r$ are spin operators at the left
(L) and right (R)-hand sites respectively on each rung $r$ of the ladder.
When $J_\|=0$,
the system consists of $N$ noninteracting dimers and
the eigenstates are direct products of singlet or triplet states
of the dimers.
To the first order in $J_\|/J_\perp$,
the lowest excited states are given by Bloch states
\begin{equation}\label{eq:Bloch}
  |k\sigma\rangle=\frac{1}{\sqrt{N}}\sum_{r=1}^{N}{\rm e}^{ikr}|s\ldots
t^\sigma_r\ldots s\rangle,
\end{equation}
where the $r$th rung is excited to a triplet with the $z$ component of spin
$\sigma$,
and have a cosine dispersion
\begin{equation}\label{eq:cos}
\epsilon_{k} = J_{\perp} + J_{\|}\cos k.
\end{equation}
Then we obtain the matrix element
\begin{eqnarray}
|\langle k',\sigma+1|S^L_{r+}|k\sigma\rangle|^2 & =& \frac{1}{N^2}|{\rm  e}
^{i(k-k')r}\langle t_r^{\sigma+1}|S^L_{r+}|t_r^\sigma\rangle|^2\nonumber\\
 &=& \frac{1}{2N^2},
\end{eqnarray}
and thus
\begin{eqnarray}
\frac{1}{T_1}&=&\frac{A_\perp^2}{4\hbar^2}\frac{1}{N^2}
\sum_{k,k'}(\textrm{e}^{-\beta\epsilon_{k}}+\textrm{e}^{-\beta(\epsilon_
{k}+h)})\nonumber\\
&&\times2\pi\hbar\,\delta\left(\frac{\epsilon_k-\epsilon_{k'}+h}{\hbar}-\omega_
{\rm N}\right),
\end{eqnarray}
where $h\equiv g\mu_BH$.
Since $h\gg\hbar\omega_{\rm N}$,
the Zeeman splitting of the magnon bands is essential and $\omega_{\rm N}$
is negligible.
Replacing the sum by integral
\[\frac{1}{N}\sum_{k}\ldots \rightarrow \int_0^\infty d\epsilon\,\rho(\epsilon)
\ldots,\]
we obtain
\begin{equation}\label{eq:T1}
 \frac{1}{T_1}= \frac{\pi A_\perp^2}{2\hbar}\int_0^\infty d\epsilon\,
\rho(\epsilon)\rho(\epsilon+h){\rm e}^{-\beta\epsilon}(1+{\rm e}^{-\beta h}),
\end{equation}
where $\rho(\epsilon)$ is the magnon density.
For the cosine dispersion eq.~(\ref{eq:cos}), we obtain
\begin{eqnarray}\label{eq:cosT1}
 \frac{1}{T_1}&=&\frac{A_\perp^2}{2\pi\hbar J_\|}
{\rm e}^{-\beta\Delta}(1+{\rm e}^{-\beta h})\nonumber\\
&&\times\int_{-1}^{1-h/J_\|}\,dx\frac{{\rm e}^{-\beta J_\|(1+x)}}
{\sqrt{1-x^2}\sqrt{1-(x+h/J_\|)^2}},\hspace{8mm}
\end{eqnarray}
where $\Delta = J_\perp-J_\|$.
If we expand the dispersion around the band bottom $k=\pi$ as
\begin{equation}
  \epsilon_{k} = \Delta + \frac{c^2}{2\Delta}(\pi -|k|)^2,
\end{equation}
where the magnon velocity $c = \sqrt{\Delta J_\|}$, we obtain
\begin{eqnarray}\label{eq:paraboT1}
 \frac{1}{T_1}&\approx &\frac{A_\perp^2}{2\pi\hbar J_\|}{\rm e}^
{-\beta\Delta}\cosh\frac{\beta h}{2}\, K_0\!\left(\frac{\beta h}{2}
\right),
\end{eqnarray}
where $K_0(x)$ is the modified Bessel function of the second kind.
Troyer {\it et al}.\ obtained a similar result\cite{Troyer} but 
its magnetic-field dependence is unrealistic because they neglected
the Zeeman splitting of the magnon bands.

Next we consider the trellis lattice including the inter-ladder coupling $J'$.
We have to treat a two dimensional magnon dispersion.
To the first order in $J_\|/J_\perp$ and $J'/J_\perp$,
the triplet magnon band splits into two branches as\cite{Miya}
\begin{equation}
\epsilon(k_\perp,k_\|) = J_{\perp}+J_{\|}\cos k_\| \pm |J'|\cos
\frac{k_\perp}{2}\cos\frac{k_\|}{2},
\end{equation}
and these states are given by Bloch states similarly to eq.~(\ref{eq:Bloch}).
Then we obtain the matrix elements
\begin{equation}
  |\langle {\mib k}',\alpha',\sigma+1|S^L_{r+}|{\mib k},\alpha,
\sigma\rangle|^2 = \frac{1}{2N^4},
\end{equation}
where ${\mib k}=(k_\perp, k_\|)$ and $\alpha$ indicates one of the two
branches.
$1/T_1$ is given by
\begin{eqnarray}
\frac{1}{T_1}&=&\frac{A_\perp^2}{4\hbar^2}\frac{1}{N^4}
\sum_{\mib k,k'}\sum_{\alpha,\alpha'}(\textrm{e}^{-\beta\epsilon({\mib k},
\alpha)}+\textrm{e}^{-\beta\epsilon({\mib k'}\!, \alpha')+h)})\nonumber\\
&&\times2\pi\hbar\,\delta\left(\frac{\epsilon({\mib k},\alpha)-\epsilon
({\mib k}'\!,\alpha')+h}{\hbar}-\omega_{\rm N}\right),
\end{eqnarray}
and using the magnon density we finally obtain the same expression as
eq.~(\ref{eq:T1}).

For both the two-leg ladder and the trellis lattice,
$1/T_1$ depends on not only  the energy gap $\Delta$ but also the whole
magnon  density,
thereby we can estimate the exchange parameters
from the analysis of $1/T_1$.

\section{Experimental Results and Discussion}

In this section we present the experimental results and compare it with
the theoretical calculation.

\subsection{Temperature Dependence of $1/T_1$}

\begin{figure}[bt]
  \begin{center}
\epsfxsize=80mm \epsfbox{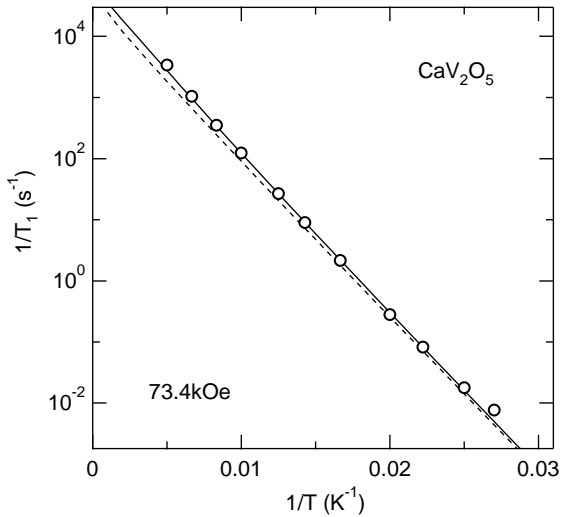}
  \end{center}
\caption{Temperature dependence of 1/$T_1$. Solid line shows a fit
to the strong-coupling expansion eq.~(\ref{eq:cosT1}) (the cosine dispersion).
Dashed line is calculated with the parabolic dispersion with the same
$J_\perp$ and $J_\|$.}
\label{fig:Tdep}
\end{figure}
Figure \ref{fig:Tdep} shows the temperature dependence of $1/T_1$
measured with the magnetic field of 73 kOe.
We observed activated temperature dependence down to 40 K.
This clearly indicates
the singlet ground state, and thus contradicts a $\mu$SR measurement
which suggests spin freezing below $\sim$ 50 K.\cite{Luke}
This is most likely an impurity effect caused by incident muons.

We first compare the observed temperature dependence of $1/T_1$
with a numerical calculation of eq.(\ref{eq:cosT1}) for the two-leg ladder
without the inter-ladder coupling.
Using $A_\perp=\sqrt{(A_x^2+A_y^2)/2}\sim 5.9\times10^{-3}$ K,\cite{OhamaCa}
we can fit the $1/T_1$ data well with $J_\perp\sim$
 655(5) K,
$J_\|\sim$ 93(5) K, and consequently $\Delta\sim$ 562(5) K.
This confirms the anisotropic exchange interactions $J_\perp \gg J_\|$.
The calculated temperature dependence is in good agreement with the data except
35 K and above 150 K as shown in Fig.~\ref{fig:Tdep}.
At the low temperatures the deviation is most likely due to magnetic
impurities and defects.
At high temperatures we expect that the independent magnon approximation is
not applicable.

\subsection{Magnetic field dependence of $1/T_1$}

\begin{figure}[tb]
  \begin{center}
\epsfxsize=85mm \epsfbox{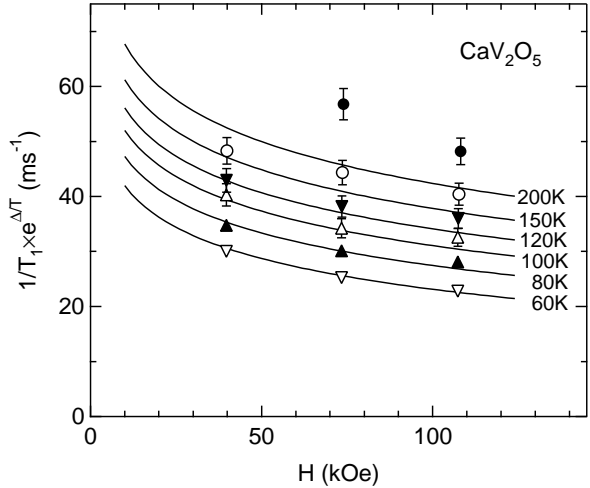}
  \end{center}
\caption{Magnetic field  dependence of $1/T_1$.
Solid line shows a fit to the free boson theory eq.~(\ref{eq:cosT1}).
We take $\Delta\sim$ 562 K.}
\label{fig:Hdep}
\end{figure}

Figure \ref{fig:Hdep}\ shows the magnetic field dependence of $1/T_1$;
we found weak but clear dependence.
We calculate the magnetic field dependence numerically with 
eq.~(\ref{eq:cosT1}) using the exchange parameters determined in the previous
section $J_\perp = 655$ K, $J_\| = 93$ K.
This calculation agrees with the experimental data
as shown in Fig.~\ref{fig:Hdep}.

Recently Sachdev and Dample developed a semiclassical continuum theory for
one-dimensional gapful antiferromagnets.\cite{Sachdev97,Damle}
They considered magnon-magnon collision and found 
a diffusive behavior in $1/T_1$.
They found that $1/T_1\propto {\rm e}^{-3\beta\Delta/2}$ not
$\propto {\rm e}^{-\beta\Delta}$ at `not too low' temperatures.
They claimed that this `large gap' 3$\Delta$/2 explains the discrepancy
between energy gaps determined from $1/T_1$ and magnetic susceptibility
in one-dimensional gapful antiferromagnets.
At low temperatures this theory agrees with the independent magnon
approximation eq.~(\ref{eq:paraboT1}) but at high temperatures or
at low magnetic fields the effect of magnon-magnon collision appears.
We compare these theories and investigate whether
the magnon-magnon collision is important in the temperature and 
magnetic field ranges of our measurement in \cavo .

They obtain the dynamical structure factor for
a parabolic dispersion $\epsilon_{k} = \Delta + (c^2/2\Delta)
(\pi -|k|)^2$,
\begin{eqnarray}
  S_{\mp,\pm}(\omega_{\rm N})&=&  
\frac{2(\rho_0+\rho_{\mp 1})}{c}
\sqrt{\frac{2\Delta}{\pi T}}\left\{\ln (L_t T)\right.\nonumber\\
&&\left.+\Phi_2(\sqrt{\pi}\,|\hbar\omega_{\rm N}\pm h|
\,L_t)\right\},
\end{eqnarray}
where
\begin{equation}
  \rho_m=\sqrt{\frac{T\Delta}{2\pi c^2}}{\rm e}^{-\beta(\Delta-mh)},
\end{equation}
and $L_t=\sqrt{\pi}\beta{\rm e}^\Delta/3T$.
The scaling function is
\begin{eqnarray}
  \Phi_2(\Omega)&=&\ln\left(\frac{4\sqrt{\pi}{\rm e}^{-\gamma}}{\Omega}\right)
+\frac{\pi\left[(\sqrt{4+\Omega^2}+2)^\frac{1}{2}-\sqrt{\Omega}\right]^2}
{4\sqrt{\Omega}\,(\sqrt{4+\Omega^2}+2)^\frac{1}{2}}\nonumber\\
&&+\ln\frac{[1+\Omega^2/\Psi^2(\Omega)]^\frac{1}{2}[1+\Psi(\Omega)]}{2\Omega},
\end{eqnarray}
\begin{equation}
  \Psi(\Omega)=\left(\Omega\sqrt{1+\frac{\Omega^2}{4}}-\frac{\Omega^2}{2}
\right)^\frac{1}{2},
\end{equation}
where $\gamma$ is Euler's constant.
For the anisotropic two-leg ladder $1/T_1$ is given by
\begin{equation}
  \frac{1}{T_1}=\frac{A_\perp^2}{4\hbar}\frac{S_\pm(\omega_{\rm N})+
S_\mp(\omega_{\rm N})}{2}.
\end{equation}

\begin{figure}[tb]
  \begin{center}
\epsfxsize=80mm \epsfbox{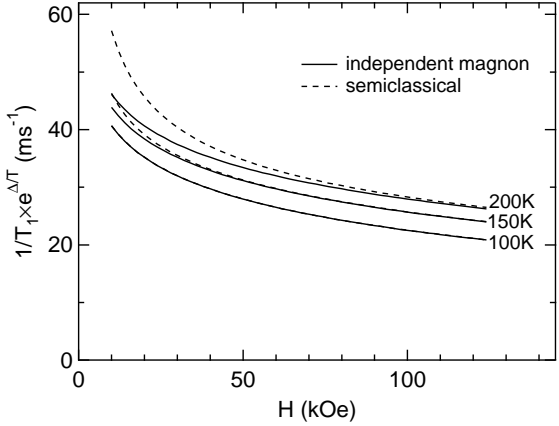}
  \end{center}
\caption{Calculated magnetic field dependence of $1/T_1$ by independent
magnon approximation and semiclassical continuum theory with
 the parabolic dispersion with
$\Delta=562$ K and $c=229$ K.}
\label{fig:diffuse}
\end{figure}
In Fig.~\ref{fig:diffuse}, we compare the theoretical results
for $\Delta=J_\perp-J_\|\sim $ 562 K and $c=\sqrt{\Delta J_\|}\sim$ 229 K.
The temperature dependence for the parabolic dispersion
with these exchange parameters is shown in Fig.~\ref{fig:Tdep}.
Below 150 K, the difference is negligible at the magnetic fields where
$1/T_1$ was measured.
At 200 K, we find a strong enhancement of $1/T_1$ due to the magnon-magnon
collision below $\sim$ 80 kOe.
From these calculations, we found that
the independent magnon approximation is good 
for the analysis of our experimental data except at 200 K.
Also we conclude that the discrepancy between the energy gap deduced from
$1/T_1$ and the NMR shift reported in the previous measurement\cite{Iwase}
cannot be explained by this semiclassical theory.

\subsection{Effects of inter-ladder coupling}

\begin{figure}[tb]
  \begin{center}
\epsfxsize=80mm \epsfbox{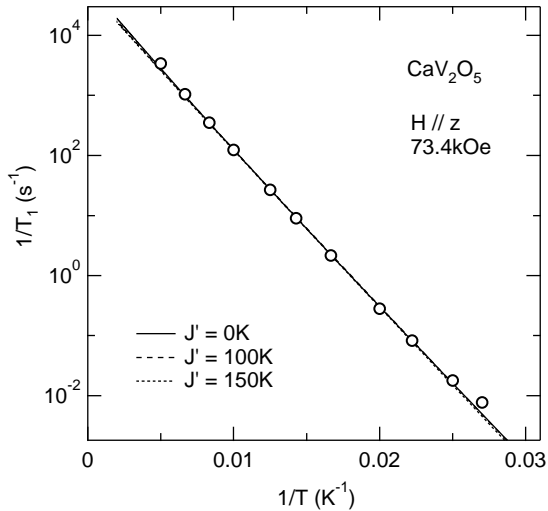}
  \end{center}
\caption{Temperature dependence of $1/T_1$ calculated with the inter-ladder
coupling $J'$.
Solid line is for $J_\perp = 655$ K, and $J_\|=$ 93 K, and $|J'|=$ 0 K,
dashed line for $J_\perp=$ 667 K, $J_\|=$ 88 K, and $|J'|=$ 100 K, and
dotted line for $J_\perp=$ 677 K, $J_\|=$ 81 K, and $|J'|=$ 150 K.
Three lines  are almost identical.}
\label{fig:trellis}
\end{figure}
We next discuss the effects of the inter-ladder coupling $J'$.
We numerically calculate $1/T_1$ with eq.~(\ref{eq:T1}) including $J'$
and obtain good fits to the data with $0\leq |J'|\ \lsim\ 150$ K.
The inter-ladder coupling affects $1/T_1$ only weakly in this range of $J'$; 
we can reproduce $1/T_1$ with similar values of $J_\perp$ and $J_\|$ 
even when $J'$ changes.
For example, $J_\perp=$ 667 K and $J_\|=$ 88 K for $|J'|=$ 100 K, and
$J_\perp=$ 677 K and $J_\|=$ 81 K for $|J'|=$ 150 K.
$1/T_1$ for these sets of exchange parameters is shown in
Fig.~\ref{fig:trellis}; the difference is very small.
Miyahara {\it et al.} calculated high temperature susceptibility
by quantum Monte Carlo simulation and obtained an estimate
$J_\perp\sim$ 670 K, $0\ \lsim\ J_\|\ \lsim$ 200 K, and $J'+J_\|\sim 110$ K.
Their estimate agrees with our analysis of $1/T_1$.
For $|J'|\ \gsim\ 150$ K, we found no range of $J_\perp$ and $J_\|$
which reproduces the data.

\subsection{Magnetic susceptibility}

Onoda and Nishiguchi reported\cite{Onoda} that the measured susceptibility
could be fitted by the dimer model with $J_\perp=664$ K
\begin{equation}
  \chi(T) = \frac{g^2\mu_{\rm B}^2}{T}
  \frac{{\rm e}^{-J_\perp/T}}{1+3{\rm e}^{-J_\perp/T}}.
\end{equation}
However, our experimental result\cite{OhamaCa} cannot be fitted
in the whole temperature
range with $g = 1.957$ measured by ESR.\cite{Onoda}
This suggests that we need to consider inter-dimer couplings $J_\|$ and $J'$.
In this section
we compare the low-temperature susceptibility with the result of the
strong-coupling expansion to check
whether it can describe the temperature dependence of
both the susceptibility and $1/T_1$ consistently.

Troyer {\it et al.} gave an expression for the susceptibility of
the two-leg ladder near the strong-coupling limit:\cite{Troyer}
\begin{equation}\label{eq:X}
\chi(T) = \frac{g^2\mu_{\rm B}^2}{T}\frac{z(\beta)}{1+3z(\beta)},
\end{equation}
\begin{eqnarray}
z(\beta)&\equiv&\frac{1}{2\pi}\int_{-\pi}^{\pi} dk\ {\rm e}^
{-\beta\epsilon_k}\\
&=&\int_{0}^{\infty} d\epsilon\ \rho(\epsilon)\,{\rm e}^{-\beta\epsilon_k}.
\label{eq:zb}
\end{eqnarray}
This expression is correct in both low and high temperature limits.
For the cosine dispersion eq.~(\ref{eq:cos}), we obtain
\begin{equation}
z(\beta) = {\rm e}^{-\beta J_{\perp}} I_{0}(\beta J_{\parallel}),
\end{equation}
where $I_0(x)$ is the modified Bessel function of the first kind
and the susceptibility is given by eq.~(\ref{eq:X}).
Similarly we can calculate the susceptibility numerically
for the trellis lattice
including the inter-ladder coupling $J'$
from eqs.~(\ref{eq:X}) and (\ref{eq:zb}) using the magnon
density $\rho(\epsilon)$.

\begin{figure}[tb]
  \begin{center}
\epsfxsize=83mm \epsfbox{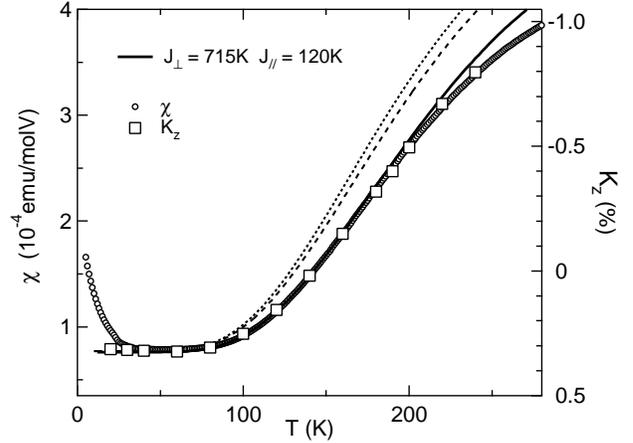}
  \end{center}
\caption{Temperature dependence of magnetic susceptibility:
solid line and dotted line are two-leg ladder models with
$J_\perp= 715$ K and $J_\|=120$K, and $J_\perp= 655$ K and $J_\|=93$K, 
respectively.
Dashed line is trellis lattice model with $J_\perp= 667$ K, $J_\|=88$ K, 
and $J'=100$ K.
Small circle and large square are experimental results of magnetic
susceptibility and NMR shift at the V sites, respectively.\cite{OhamaCa}}
\label{fig:XvsT}
\end{figure}
In Fig.~\ref{fig:XvsT} we show the experimental susceptibility.\cite{OhamaCa}
To compare it to the theory, we have to obtain the spin susceptibility
by subtracting the temperature-independent offset of the orbital and
diamagnetic susceptibilities and
magnetic impurity contributions (Curie tail).
Since the experimental susceptibility is almost temperature-independent
between 30 and 70 K,
the susceptibility in this temperature range
is a possible estimate of the sum of the orbital and diamagnetic
susceptibilities.
Then the increase below 30 K in decreasing temperature is attributed
to impurity contributions.
Since NMR shift is insensitive to such impurities,
we can check if the observed temperature dependence of the susceptibility
is intrinsic by comparing the susceptibility  and the NMR shift.
We plot the NMR shift at V sites\cite{OhamaCa} in Fig.~\ref{fig:XvsT}.
It is found that the NMR shift fits to the susceptibility well down to 40 K
and that it is temperature-independent down to 20 K.
From this fitting we found that the observed temperature dependence
above $\sim$ 50 K is intrinsic of the spin susceptibility
and that the Curie tail is negligible above about 40 K.
We thereby obtain an estimate of the sum of the orbital and diamagnetic
susceptibilities $\sim 7.7\times 10^{-5}$ emu/molV.
Although the $K$-$\chi$ plot indicates that this `residual susceptibility'
is too large,\cite{OhamaCa,remark}
we use this estimate in the following analysis since we have no other reliable
estimates.

We first  calculate the susceptibility for the exchange parameters
obtained from the analysis of $1/T_1$;
for the two-leg ladder model  $J_\perp= 655$ K and $J_\|=93$K, 
and for the trellis lattice model  $J_\perp= 667$ K, $J_\|=88$ K, 
and $J'=100$ K.
These parameters cannot reproduce the experimental susceptibility
as shown in Fig.~\ref{fig:XvsT}.
Instead we can reproduce the low temperature susceptibility below 200 K
with $J_\perp =$ 715 K and $J_\| =$ 120 K for the two-leg ladder model.
The inter-ladder coupling $J'$ affects the susceptibility only weakly 
and does not eliminate the disagreement between the exchange parameters
from $1/T_1$ and the susceptibility:
for example we obtain  almost the same temperature dependence with 
the parameters $J_\perp =$ 720 K, $J_\| =$ 120 K, $J' =$ 100 K.
These results suggest the energy gap of $\Delta \sim$ 600 K.
The previous result of the NMR shift $\Delta \sim 460$ K is
an underestimate due to the large experimental error in the NMR shift
measurement.\cite{Iwase}

We thus found a discrepancy between the exchange parameters from 
$1/T_1$ and the susceptibility.
This may be reduced by i) a $1/T_1$ calculation with higher-order of 
$J_\|/J_\perp$, ii) consideration of the off-diagonal part of the hyperfine
interaction which was neglected, iii) a reliable estimate of the spin
susceptibility, but we have no convincing explanations at present.

\section{Conclusions}

We measured $1/T_1$ at the V sites in the two-leg ladder
antiferromagnet CaV$_2$O$_{5}$.
We calculated $1/T_1$ for the two-leg ladder with the strong-coupling expansion
($J_\perp \gg J_\|$) within an independent magnon approximation,
and compared the result with the experimental data.
The comparison of the temperature dependence shows a good agreement and
provides an estimate of both $J_\perp\sim 655(5)$ K and $J_\|\sim 93(5)$ K
as well as the energy gap $\Delta\sim 562(5)$ K;
the anisotropic exchange interactions $J_\perp \gg J_\|$ was established.
The magnetic field dependence of $1/T_1$ is also reproduced within this
approximation below 150 K $\sim\Delta/4$.
It is found that the magnon-magnon collision is negligible
in this temperature range.
We also studied the effect of inter-ladder coupling $J'$.
It is found that $J'$ influences $1/T_1$ only weakly and is in the range
$0 \leq |J'|\ \lsim\ 150$ K.

We also calculated the magnetic susceptibility and compared with the
previous experimental data. 
A discrepancy between the exchange parameters deduced from the magnetic
susceptibility and $1/T_1$ is found,
although $1/T_1$ can be satisfactorily understood within our calculation
with the strong-coupling expansion.

\section*{Acknowledgments}
We would like to thank Y.~Itoh, J.~Kikuchi, and H.~Fukumoto for helpful
discussions, and H.~Tsunetsugu for useful comments.

\end{document}